\documentclass[%
 reprint,
 amsmath,amssymb,
 aps,
]{revtex4-1}

\usepackage{graphicx}% Include figure files
\usepackage{dcolumn}% Align table columns on decimal point
\usepackage{bm}% bold math
\usepackage{algorithm}
\usepackage{algorithmic}
\usepackage{subfigure}

\begin{document}

\preprint{APS/123-QED}

\title{Bayesian Modeling of Random Walker for Community Detection in Networks}% Force line breaks with \\
%% \thanks{A footnote to the article title}%

\author{Takafumi J. Suzuki}
 \email{suzuki.taka-fumi@fujixerox.co.jp}
\affiliation{
  Communication Technology Laboratory, Research $\&$ Technology Group, Fuji Xerox Co., Ltd.\\
  6-1 Minatomirai, Nishi-ku, Yokohama, Kanagawa, 220-8668, Japan
}
\date{\today}

\begin{abstract}
  We propose a generative model to detect globally optimal community structures in networks by utilizing random walks.
  %% We develop a stochastic method to detect community structures in networks by extending an existing model for detecting link communities with random walks.
  Sophisticated parameter optimization algorithms are developed based on the Markov chain Monte Carlo methods to overcome limitations of the EM algorithm, which has been used in previous works but is sometimes trapped in local optima depending on initial conditions.
  We apply the algorithms to synthetic and real-world networks to examine their performance in terms of precision and robustness of detected communities.
  It is found that the Gibbs samplers outperform the previous approaches especially in detecting overlapping communities.
  The Markovian dynamics of random walkers is crucial to robustly detect the optimal community structures.
  %% We further analyze optimization processes of community assignments and demonstrate that the Markovian dynamics of random walkers is crucial to robustly detect optimal community structures.
\end{abstract}

%\keywords{Suggested keywords}%Use showkeys class option if keyword
                              %display desired
\maketitle

%\tableofcontents

\section{Introduction}

Network structures are ubiquitously found in a wide variety of fields in science such as biology, sociology, and computer science~\cite{barabasi2016network}.
Nodes and links in networks represent fundamental elements and their interactions in the target systems.
These constituents are often spontaneously organizing into structures called {\it communities}, i.e. groups of nodes which are densely connected to each other but sparsely connected to the rest of the network.
In the last few decades, a large number of methods have been developed to detect communities because uncovering the underlying community structures is an essential step to clarify both microscopic and macroscopic functionalities of the networks~\cite{Girvan7821,newman2006modularity,PhysRevE.80.056117,FORTUNATO201075,ball2011efficient,Newman2011,FORTUNATO20161}.

Among the various approaches for community detection, stochastic modeling of generative processes in networks has been recognized as a promising technique, and formed the active field of research~\cite{ball2011efficient}.
The stochastic block model and its variants are prominent examples of such models~\cite{holland1983stochastic,nowicki2001estimation,PhysRevE.83.016107,JMLR:v18:16-480}.
These models are based on an idea that links in the networks are generated with probability determined by community assignments of their endpoint nodes:
The nodes belonging to the same community are more likely to be connected than those in different communities.
This simple idea can be flexibly extended to incorporate various aspects of the networks such as
heterogeneity in the degrees of node~\cite{PhysRevE.83.016107}
and
their mixed memberships~\cite{airoldi2008mixed}.
%infinite relational model~\cite{kemp2006learning}
Another successful approach for detecting communities is to utilize random walkers on networks.
Random walks are stochastic processes of the agents who randomly select their paths and travel around the networks.
The random walks have formed the basis of modern community-detection methods such as Walktrap~\cite{pons2005computing} and Infomap~\cite{rosvall2007information,rosvall2009map} because they offer intuitive and useful pictures of elementary dynamics of the networks.

Recent research has shown that fundamental characteristics of real-world networks can be reasonably explained by communities of {\it links} rather than nodes~\cite{ball2011efficient}.
In particular, overlapping of node communities is understood as a natural consequence of link communities~\cite{PhysRevE.80.016105,ahn_link_2010}.
The key point is that most of the links in typical networks have unique properties to characterize relationships between the nodes.
For instance, links in social networks have clear meanings such as kinship or coworker relationships, while each individual plays multiple roles in the concerned networks.
These observations have promoted the development of methods for detecting link communities~\cite{ahn_link_2010,Xie:2013:OCD:2501654.2501657,ding_overlapping_2016,PhysRevE.80.016105,He:2015:SMD:2887007.2887026,ball2011efficient,He2015}.

%% link communities~\cite{ahn_link_2010,He:2015:SMD:2887007.2887026}
%% generative network model proposed by B. Ball {\it et. al}~\cite{ball2011efficient},
%% link model with iterative bipartition (LMBP)~\cite{He:2015:SMD:2887007.2887026}
%% hybrid node-link communities~\cite{He2015}

% Modularity based approach~\cite{newman2006modularity}.

%% %% sentence from ``A Stochastic Model for Detecting Heterogeneous Link Communities in Complex Networks''
%% In many real networks, link communities are often more informative and
%% intuitive than node communities, because links usually
%% have unique identities, while nodes typically have multiple
%% roles. In a social network, for instance, most individuals
%% belong to multiple communities such as families, friends,
%% and co-workers, while the link between two individuals
%% often exists for a dominant reason which may represent
%% family ties, friendship, or professional relationships. Furthermore, multiple links connecting to a node may belong
%% to distinct link communities, so that the node can be assigned to multiple communities of links. Accordingly,
%% overlapping communities of nodes, another attractive topic
%% in community detection (Palla et al. 2005), could be detected as a natural byproduct of link communities. 

Modular decomposition of Markov chain (MDMC) is a community detection method proposed in Ref.~\onlinecite{mdmc_original} by combining the aforementioned insights into community structures.
MDMC shares common ideas with previous methods for link communities in that generative processes of links are modeled with underlying community structures.
The important difference is that random walks are incorporated in the generative processes in MDMC, while only the local pairs of nodes are considered in the previous methods.
Since the random walks are useful to capture global structures of the networks, they allow us to find optimal community structures
without being trapped in meta-stable local ones.
% Random walker approach. info map.
In return for stable and high-quality results in MDMC, however, it is necessary to solve coupled master equations, which require sophisticated schemes to optimize model parameters.
Unfortunately, the EM algorithm, which was utilized in the original formulation~\cite{mdmc_original,Okamoto2019ModularDO},
does not always find the optimal structures because it is sensitive to initial values of the parameters.
Hence, elaboration of the optimization algorithm is highly desired to make full use of the powerful modeling of MDMC.

In this paper, we develop efficient parameter optimization algorithms for MDMC based on the variational Bayesian and Markov chain Monte Carlo methods to enhance the performance of the model.
The developed algorithms are applied to both synthetic and real-world networks, and evaluated in terms of the normalized mutual information and the number of detected communities.
It is shown that Gibbs samplers enable us to robustly detect optimal community structures in a wide variety of networks compared with previous deterministic methods.
We further analyze intermediate optimization processes of the community assignments and show that the Markovian dynamics of random walkers is essential for finding the globally optimal solutions.

This paper is organized as follows.
In Sec.~\ref{sec: methodology}, we describe guiding principles of MDMC, and formulate generative processes of links with random walkers on networks.
Gibbs samplers and variational Bayesian methods for inferring the model parameters are developed based on the Bayesian formulation of MDMC.
The performance of the developed optimization algorithms is evaluated in Sec.~\ref{sec: evaluation} with synthetic and real-world networks.
We will also show optimization processes of community assignments and model parameters to understand the roles of the random walkers in the inference steps.
Concluding remarks are stated in Sec.~\ref{sec: conclusions}.

\section{Methodology}
\label{sec: methodology}

\subsection{Bayesian Formulation of MDMC}
\label{subsec: Bayesian Formulation of MDMC}

The key idea of modular decomposition of Markov chain (MDMC) is to introduce a random walker on a network and trace the Markovian dynamics of the agent in terms of the probability distributions.
With $K$ latent communities behind the network, the probability for the agent to be found at node $n$ at time $t$ is denoted by $p^{(t)}(n|k)$ provided that he is in community $k$.

Suppose that the agent is observed to be traveling on a link $d$, which connects nodes $i$ and $j$ in the network.
This information can be encoded into an $N$-dimensional vector $\tau^{(t)}_{:d}$ as
\begin{eqnarray}
\tau^{(t)}_{nd}=\left\{ \begin{array}{ll}
1 & n \in\{i,j\} \\
0 & \text{otherwise} \\
\end{array} \right. .
\end{eqnarray}
Hereafter, the colon is used as a placeholder of the corresponding arguments.
We repeat this procedure for every observable link $d=1$, $2$, $\cdots$, $D$ in the network to get observed data $\tau^{(t)}$ on the positions of the agent.
In order to identify community assignments of the links, we introduce a $K$-dimensional one-hot vector $z^{(t)}_{d}$ for each link $d$:
When the link $d$ belongs to community $k$, the $k$-th component of the latent variable is set as $z^{(t)}_{dk} = 1$.
Mixtures of the link communities are described by an additional $K$-dimensional parameter $\pi^{(t)}_{d}$, which satisfies the condition $\sum^{K}_{k=1}\pi^{(t)}_{dk}=1$.
%% We assume that the mixture parameter $\pi^{(t)}_{d}$ is also sampled from another distribution with parameter $\eta^{(t)}_{k}$.
%% Although we have considered that observation is performed on the link, it is straightforward to generalize the discussions into the observation of cliques.
When the link $d$ connecting nodes $m$ and $n$ is assigned to, let's say, community $k$, the probability for the link to be generated is assumed to be proportional to $p^{(t)}(n|k)p^{(t)}(m|k)$.
The probability of observing the data $\tau^{(t)}_{:d}$ is obtained by summing the above probabilities over all the possible communities.

\begin{figure}[htbp]
    \includegraphics[width=3.0cm]{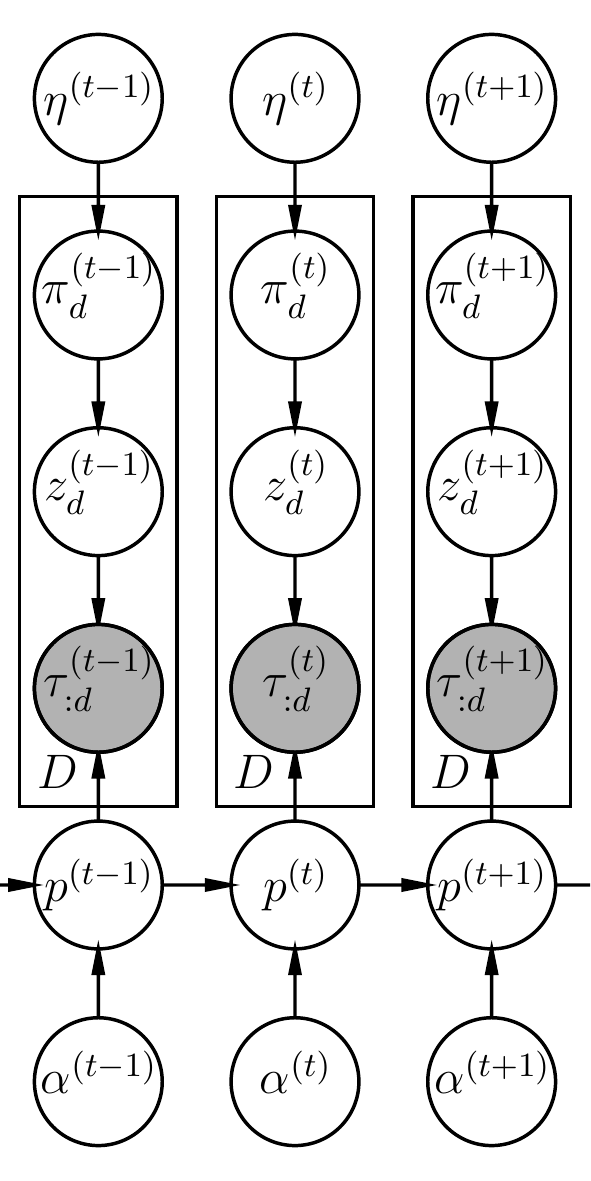}
    \caption{Graphical model representation of MDMC}
    \label{fig: graphical representation}
\end{figure}

\renewcommand{\arraystretch}{1.1}
\begin{table*}
  \caption{
    \label{table: MDMC notations}
    Notations of parameters and variables in MDMC
  }  
  \begin{ruledtabular}
    \begin{tabular}{lll}
      \textrm{Notation}&
      \textrm{Description}&
      \textrm{Dimension}\\
      \colrule
      $N$ & \textrm{Number of nodes} & 1\\
      $D$ & \textrm{Number of links} & 1\\
      $K$ & \textrm{Number of communities} & 1\\
      $T_{\rm step}$ & \textrm{Number of Markovian time steps} & 1      \\
      $S$ & \textrm{Number of Monte Carlo sampling} & 1      \\
      $T_{\rm iter}$ & \textrm{Number of iterations in deterministic algorithms} & 1 \\
      $t$ & \textrm{Time of Markov chain} & 1      \\
      $\alpha^{(t)}_{k}$ & \textrm{Parameters of Dirichlet distribution for} $p^{(t)}(:|k)$ & $K$ \\
      $\eta^{(t)}_{k}$ & \textrm{Parameters of Dirichlet distribution for} $\pi^{(t)}$ & $K$ \\
      $T_{nm}$ & \textrm{Transition matrix} & $N \times N$ \\
      $\tau^{(t)}_{nd}$ & \textrm{Observed data} & $N \times D$ \\
      $z^{(t)}_{dk}$ & \textrm{Latent variables for community assignments} & $D \times K$ \\
      $p^{(t)}(n|k)$ & \textrm{Probability distributions of random walkers} & $N \times K$ \\
      $\pi^{(t)}_{dk}$ & \textrm{Mixture parameters} & $D \times K$
    \end{tabular}
  \end{ruledtabular}
\end{table*}

In the Bayesian formulation, MDMC consists of the following generating processes at time $t$.
\begin{enumerate}
\item A probability distribution $p^{(t)}$ is sampled from the Dirichlet prior distribution;
\begin{align}
  \label{eq: pt}
  &P(p^{(t)}|\alpha^{(t)}) = \prod^{K}_{k=1} \text{Dir}(p^{(t)}(:|k)|\alpha^{(t)}(:|k)),
\end{align}
where $\text{Dir}(x|\beta)$ denotes the Dirichlet distribution
\begin{align}
  \label{eq: Dirichlet distribution}
  \text{Dir}(x|\beta)
  &= 
  \frac{\Gamma\left( \sum^{N}_{n=1} \beta_{n} \right)}
       {\prod^{N}_{n=1} \Gamma(\beta_{n})}
       \prod^{N}_{n=1} \left( x_{n} \right)^{\beta_{n}-1},
\end{align}
with the gamma function $\Gamma(x)$.
The parameter of the distribution (\ref{eq: pt}) is defined by
\begin{equation}
  \label{eq: definition of alpha}
  \alpha^{(t)}(n|k) := \alpha^{(t)}_{k} \sum^{N}_{m=1} T_{nm}p^{(t-1)}(m|k)+1,
\end{equation}
with the transition matrix $T_{nm}$ ($n,m=1,2,\cdots,N$), which satisfies $\sum^{N}_{n=1}T_{nm}=1$.
We note that the probability distribution at time $t$ depends only on the latest one $p^{(t-1)}$.
This assumption leads to Markovian dynamics of random walkers.
It is shown later that the maximum a posteriori (MAP) estimation of the parameter $p^{(t)}$ results in master equations for a network augmented by the observed data $\tau^{(t)}$ [see Eq.~(\ref{eq: p_t point esimate}) in Sec.~\ref{sec: inference methods}].
\item For $d=1,2,\cdots,D$, a $K$-dimensional mixture parameter $\pi^{(t)}_{d}$ is independently sampled from the Dirichlet distribution with a prior parameter $\eta^{(t)}$ of the same dimension;
  \begin{align}
    \label{eq: pi_t}
    P(\pi^{(t)}_{d}|\eta^{(t)})
    &=  \text{Dir}(\pi^{(t)}_{d}|\eta^{(t)}).
  \end{align}
\item A latent variable $z^{(t)}_{d}$ for community assignments of the link $d$ is sampled from the categorical distribution with the mixture parameter $\pi^{(t)}_{d}$;
\begin{align}
  \label{eq: pz}
  P(z^{(t)}_{d}|\pi^{(t)}_{d}) = \prod^{K}_{k=1}\left(\pi^{(t)}_{dk}\right)^{z^{(t)}_{dk}}.
  %% P(z^{(t)}_{dk}=1|\pi^{(t)}_{d}) = \pi^{(t)}_{dk}.
\end{align}
\item Given that the link $d$ belongs to community $k$, the observed data $\tau^{(t)}_{:d}$ obeys the multinomial distribution;
\begin{align}
  \label{eq: tau}
  &P(\tau^{(t)}_{:d}|z^{(t)}_{dk}=1,p^{(t)})
  = \text{Mult}\left( \tau^{(t)}_{:d}|p^{(t)}(:|k)\right),
\end{align}
with
\begin{align}
  \label{eq: multinomial distribution}
  \text{Mult}\left( x|p \right)
  = \frac{\left(\sum^{N}_{n=1}x_n\right)! }{\prod^{N}_{n=1} x_{n}!}
  \prod^{N}_{n=1} \left( p_{n} \right)^{x_{n}}.
\end{align}
\end{enumerate}
% Here, auxiliary time index $t-1$ and $t$ denote the previous and current EM step, respectively.
The generative model composed of Eq.~(\ref{eq: pt}), (\ref{eq: pi_t}), (\ref{eq: pz}), and (\ref{eq: tau}) is graphically represented in Fig.\ref{fig: graphical representation}~\cite{bishop2006pattern}.
The grayed circles describe the observed data, while the open ones denote the random variables and the parameters to be determined in subsequent inference steps.
The rectangle plates represent that the stochastic variables inside are independently sampled for $D$ times.
Table~\ref{table: MDMC notations} lists the notations of the variables and the parameters used in MDMC.

%% The generative process of time slice $t$ is given as follows:
%% \begin{enumerate}
%% \item For each $k$, draw the distribution $p^{(t)}(:|k) \sim \text{Dir}(p^{(t)}(:|k)|\alpha(:|k))$.
%% \item For each data $d$:
%%   \begin{enumerate}
%%   \item Draw $z^{(t)}_{d:} \sim \text{Mult}\left( z^{(t)}_{d:}|\pi^{(t)}\right)$
%%   \item Draw $\tau^{(t)}_{nd} \sim \text{Mult}\left( \tau^{(d)}|p^{(t)}(:|k)\right)$ 
%%   \end{enumerate}
%% \end{enumerate}

\subsection{Approximation of Likelihood Function}
\label{subsec: approximation of likelihood function}

With the assumption that we have already obtained the parameters at previous times,
the likelihood function at time $t$ is given by
\begin{align}
  &P(\tau^{(t)} | \alpha^{(t)},\eta^{(t)}) \nonumber \\
  \label{eq: likelihood function}
  &= \sum_{z^{(t)}} \int dp^{(t)} d\pi^{(t)}
  P(\tau^{(t)}, z^{(t)},p^{(t)}, \pi^{(t)}| \alpha^{(t)},\eta^{(t)}),
\end{align}
with the joint probability distribution
\begin{align}
  &P(\tau^{(t)},z^{(t)},p^{(t)},\pi^{(t)} | \alpha^{(t)},\eta^{(t)}) \nonumber \\
  %% &= P(\tau^{(t)}|z^{(t)},p^{(t)}) P(z^{(t)}|\pi^{(t)})
  %% P(p^{(t)}|\alpha^{(t)}) P(\pi^{(t)}|\eta^{(t)}), 
  %% \nonumber \\
  &=
  \left[
    \prod^{K}_{k=1} \prod^{D}_{d=1}
    \left[  
      \pi^{(t)}_{dk}
      \text{Mult}\left( \tau^{(t)}_{:d}|p^{(t)}(:|k)\right)
      \right]^{z^{(t)}_{dk}}      
      \right]
  \nonumber \\
  \label{eq: joint probability distribution}
  & \hspace{10pt} \times
  \left[
    \prod^{K}_{k=1} \text{Dir}(p^{(t)}(:|k)|\alpha^{(t)}(:|k))
    \right]
  \left[
    \prod^{D}_{d=1}\text{Dir}\left(\pi^{(t)}_{d}|\eta^{(t)}\right)
    \right].
\end{align}

The total likelihood function, which should be optimized to infer the model parameters, is the product of the one for every $t$;
\begin{equation}
  \label{eq: Total likelihood function}
  P( \{\tau^{(t)}\}_{t=1,2,\cdots,T_{\rm step}} )
  = \prod^{T_{\rm step}}_{t=1} P(\tau^{(t)} | \alpha^{(t)},\eta^{(t)}),
\end{equation}
where the number of the whole time steps is denoted by $T_{\rm step}$.
However, direct optimization of the total likelihood function (\ref{eq: Total likelihood function}) is challenging because it requires us to simultaneously determine the parameters at the whole time steps.
In our formulation, we sidestep this issue by approximating the total likelihood function (\ref{eq: Total likelihood function}) as the product of the likelihood function (\ref{eq: likelihood function}) with parameters separately optimized for each time slice $t$:
\begin{equation}
  \label{eq: Estimate likelihood function}
  P( \{\tau^{(t)}\}_{t=1,2,\cdots,T_{\rm step}} )
  \approx \prod^{T_{\rm step}}_{t=1} P(\tau^{(t)} | \hat{\alpha}^{(t)}, \hat{\eta}^{(t)}).
\end{equation}
Here, we determine the parameters $\hat{\alpha}^{(t)}$ and $\hat{\eta}^{(t)}$ by using the {\it estimate} of the probability distribution $\hat{p}^{(t-1)}$ obtained at the previous time step.
Thus, we cut the dependency chain in Fig.~\ref{fig: graphical representation}, and the problem boils down to the sequential optimization of Eq.~(\ref{eq: likelihood function}) with given $\hat{p}^{(t-1)}$.
This strategy is similar to the one utilized in the topic tracking model~\cite{Iwata:2009:TTM:1661445.1661674}.
In following discussions, we omit the hat of the estimated parameters because they may cause no confusion.
% In the following discussions, we denote both the random variable and its estimate by $p^{(t)}$ because they may cause no confusion.
% We note that this formulation can naturally describe the dynamics of the network by incorporating it into the transition matrix $T^{(t)}_{mn}$ and the observed data $\tau^{(t)}_{nd}$.

\subsection{Inference Methods of MDMC Parameters}
\label{sec: inference methods}

In this subsection, we develop stochastic [Gibbs and collapsed Gibbs sampling] and deterministic [variational Bayesian and EM algorithms] methods to determine the stochastic variables $p^{(t)}$, $\pi^{(t)}$, and $z^{(t)}$ in MDMC.
At the same time, we will derive update equations of the hyperparameters $\alpha^{(t)}_{k}$ and $\eta^{(t)}_{k}$ by approximately maximizing the likelihood function (\ref{eq: likelihood function}).

\subsubsection{Gibbs Sampling}

We will first develop the Gibbs sampling algorithm, where the stochastic variables $p^{(t)}$, $\pi^{(t)}$, and $z^{(t)}$ are numerically sampled from their posterior distributions.
The probability distribution of $p^{(t)}(:|k)$ on the condition that the other parameters are fixed is calculated as
\begin{align}
  &P(p^{(t)}(:|k)|\tau^{(t)}, z^{(t)},p^{(t)}(:|\backslash k),\pi^{(t)},\alpha^{(t)},\eta^{(t)}) \nonumber \\
  \label{eq: p_t probability}
  &= \text{Dir}(p^{(t)}(:|k)|\alpha^{(t)}(:|k)+(\tau z)^{(t)}_{:k}),
\end{align}
with $(\tau z)^{(t)}_{nk} = \sum^{D}_{d=1}\tau^{(t)}_{nd}z^{(t)}_{dk}$.
Hereafter, the backslash symbol $\backslash x$ is used to denote a set of the components other than $x$.
Similarly, the probability distribution of the mixture parameter $\pi^{(t)}_{d}$ can be obtained as
\begin{align}
  \label{eq: pi_t probability}
  &P(\pi^{(t)}_{d}|\tau^{(t)},z^{(t)},p^{(t)},\pi^{(t)}_{\backslash d},\alpha^{(t)},\eta^{(t)}) \nonumber \\
  &= \text{Dir}(\pi^{(t)}_{d}|\eta^{(t)}+z^{(t)}_{d}),
\end{align}
for each $d=1$, $2$, $\cdots$, $D$.
Given the current samples of $p^{(t)}$ and $\pi^{(t)}$, the latent one-hot vector $z^{(t)}_{d}$ obeys the following probability distribution function;
\begin{align}
  \label{eq: latent variable probability}
  &P(z^{(t)}_{dk}=1|\tau^{(t)},z^{(t)}_{\backslash d},p^{(t)},\pi^{(t)},\alpha^{(t)},\eta^{(t)})
  \nonumber \\ 
  & = \frac{\pi^{(t)}_{k} \text{Mult}\left( \tau^{(t)}_{:d}|p^{(t)}(:|k)\right)}{\sum^{K}_{k=1}\pi^{(t)}_{k} \text{Mult}\left( \tau^{(t)}_{:d}|p^{(t)}(:|k)\right)}.
\end{align}
The detailed derivation of Eqs.~(\ref{eq: p_t probability}), (\ref{eq: pi_t probability}), and (\ref{eq: latent variable probability}) are given in App.~\ref{app: Gibbs Sampling}.

After taking $S$ samples of the probability distribution, we estimate the expectation value of $p^{(t)}(n|k)$ for each time slice as
\begin{align}
  \label{eq: Estimate p_t Gibbs}
  &p^{(t)}(n|k) \approx \frac{1}{S} \sum^{S}_{s=1} p^{(t)}_{;s}(n|k).
\end{align}
Hereafter, $X_{;s}$ denotes each realization of the random variable $X$ at sample $s$.

In this paper, we update the hyperparameters $\alpha^{(t)}_{k}$ and $\eta^{(t)}_{k}$ at the ends of each Markov step by approximately maximizing the likelihood function with Newton's method and Minka's fixed-point iteration~\cite{Minka2000}, respectively.
First, the parameter $\alpha^{(t)}_{k}$ is updated as
\begin{align}
  \label{eq: update of alpha}
  \alpha^{(t+1)}_{k} = \alpha^{(t)}_{k} - \frac{F_{k}(\alpha^{(t)}_{k})}{F'_{k}(\alpha^{(t)}_{k})},
\end{align}
with the logarithmic derivative of the likelihood function
\begin{align}
  F_{k}(\alpha^{(t)}_{k})
  &= \frac{d}{d\alpha^{(t)}_{k}} \ln P(\tau^{(t)}|z^{(t)},\alpha^{(t)},\eta^{(t)}) \nonumber \\
  \label{eq: F_alpha}
  &= 
    \sum^{N}_{n=1}
    \sum^{\left(\tau z\right)^{(t)}_{nk}}_{l=1}
    \frac{\left(Tp^{(t-1)}\right)_{nk}}{\alpha^{(t)}(n|k)+l-1} \nonumber \\
    & \hspace{20pt}
    - \sum^{\sum^{N}_{n=1}\left(\tau z\right)^{(t)}_{nk}}_{l=1}\frac{1}{\alpha^{(t)}_{k}+N+l-1}
  .
\end{align}
The derivation of Eq.~(\ref{eq: F_alpha}) is given in App.~\ref{app: Collapsed Gibbs Sampling}.
The updated parameters $\alpha^{(t+1)}_{k}$ and the estimate (\ref{eq: Estimate p_t Gibbs}) of the probability distribution are used to compute $\alpha^{(t+1)}(n|k)$ for the prior probability distribution of $p^{(t+1)}$ at the next time [see Eq.~(\ref{eq: definition of alpha}) for the definition of $\alpha^{(t)}(n|k)$].
%% The parameters $\alpha^{(t+1)}(n|k)$ defined in Eq.~(\ref{eq: definition of alpha}) for the probability distribution $p^{(t+1)}$ at the next time slice are computed with the updated parameters $\alpha^{(t+1)}_{k}$ and the estimated probability (\ref{eq: Estimate p_t Gibbs}).
%[see Eq.~(\ref{eq: definition of alpha}) for the definition of $\alpha^{(t)}(n|k)$].
Second, the parameter $\eta^{(t+1)}_{k}$ for the next time step is obtained by using Minka's fixed-point iteration as
\begin{align}
  \label{eq: eta inference}
  &\eta^{(t+1)}_{k}
  =  \frac{Z^{(t)}_{k}}{D}\sum^{K}_{k=1} \eta^{(t)}_{k},
\end{align}
with $Z^{(t)}_{k} = \sum^{D}_{d=1}z^{(t)}_{dk}$.
We note that the sum of the parameter $\eta^{(t)}_{k}$ is conserved during the Markov time steps because of the relation $\sum^{K}_{k=1} Z^{(t)}_{k} = D$.
%% \begin{align}
%%   \label{eq: eta inference}
%%   &\eta^{(t+1)}_{k}
%%   = \frac{\left[\Psi\left(\eta^{(t)}_{k}+Z^{(t)}_{k}\right) - \Psi\left(\eta^{(t)}_{k}\right) \right] \eta^{(t)}_{k}}{\Psi\left(\sum^{K}_{k=1} \eta^{(t)}_{k}+D\right) - \Psi\left(\sum^{K}_{k=1}\eta^{(t)}_{k}\right)}, 
%% \end{align}

To summarize, pseudocode of the Gibbs sampling is shown in Algorithm~\ref{alg: Gibbs Sampling}.
As is discussed below, we can elaborate the Gibbs sampling algorithm by analytically integrating the intermediate variables $p^{(t)}$ and $\pi^{(t)}$.
Hence, this algorithm is primarily used to check consistency of results obtained by the collapsed Gibbs sampling.

\begin{algorithm}[H]
  \caption{Gibbs Sampling}         
  \label{alg: Gibbs Sampling}
  \begin{algorithmic}[1]    
    \STATE Initialize $p^{(0)}$, $\alpha^{(1)}$, and $\eta^{(1)}$
    \FOR{$t=1,2,\cdots,T_{\rm step}$}
    \STATE Initialize $z^{(t)}$, $p^{(t)}$ and $\pi^{(t)}$
    \FOR{$s=1,2,\cdots,S$}
    \FOR{$k=1,2,\cdots,K$}
    \STATE Sample $p^{(t)}(:|k)$ [Eq.~(\ref{eq: p_t probability})]
    \ENDFOR
    \FOR{$d=1,2,\cdots,D$}
    \STATE Sample $\pi^{(t)}_{d}$ [Eq.~(\ref{eq: pi_t probability})]
    \STATE Sample $z^{(t)}_{d}$ [Eq.~(\ref{eq: latent variable probability})]
    \ENDFOR
    \ENDFOR
    \STATE Estimate $p^{(t)}$ [Eq.~(\ref{eq: Estimate p_t Gibbs})]
    \STATE Compute $\alpha^{(t+1)}$ [Eq.~(\ref{eq: update of alpha})]
    \STATE Compute $\eta^{(t+1)}$ [Eq.~(\ref{eq: eta inference})]
    \ENDFOR
  \end{algorithmic}
\end{algorithm}

\subsubsection{Collapsed Gibbs Sampling}

It is possible to analytically integrate out the stochastic variables $\pi^{(t)}$ and $p^{(t)}$ in the likelihood function (\ref{eq: likelihood function}) to get the marginalized distribution.
This property enables us to efficiently sample the latent variable $z^{(t)}$ compared with the Gibbs sampling algorithm.

The marginalized distribution is given by
\begin{widetext}
\begin{align}
  \label{eq: Marginalized likelihood function CGS}
  & P(\tau^{(t)},z^{(t)}|\alpha^{(t)},\eta^{(t)})
  &\nonumber \\
  &=
  \prod^{D}_{d=1}
  \frac{\Gamma\left( \sum^{K}_{k=1} \eta^{(t)}_{k} \right)}{\prod^{K}_{k=1} \Gamma\left(\eta^{(t)}_{k}\right)}
  \frac{\prod^{K}_{k=1}\Gamma\left( \eta^{(t)}_{k}+z^{(t)}_{dk}\right)}{\Gamma\left(\sum^{K}_{k=1} \left( \eta^{(t)}_{k}+ z^{(t)}_{dk}\right)\right)} %\nonumber \\
  %& \hspace{20pt} \times
  \prod^{K}_{k=1}
  \frac{\Gamma\left( \sum^{N}_{n=1} \alpha^{(t)}(n|k) \right)}{\prod^{N}_{n=1} \Gamma\left(\alpha^{(t)}(n|k)\right)}
  \frac{\prod^{N}_{n=1}\Gamma\left( \alpha^{(t)}(n|k)+ (\tau z)^{(t)}_{nk}\right)}{\Gamma\left(\sum^{N}_{n=1} \left( \alpha^{(t)}(n|k)+ (\tau z)^{(t)}_{nk}\right)\right)},
\end{align}
which leads to the conditional distribution of the latent variables $z^{(t)}_{dk}$;
\begin{align}
  \label{eq: z Collapsed Gibbs Sampling}
  &P(z^{(t)}_{dk}=1|\tau^{(t)},z^{(t)}_{\backslash d},\alpha^{(t)},\eta^{(t)})
  = \frac{\eta^{(t)}_{k}}{\sum^{K}_{k=1}\eta^{(t)}_{k}}
  \frac{ \prod_{n,\tau^{(t)}_{nd}\neq 0} \left(\alpha^{(t)}(n|k)+(\tau z)^{(t)}_{nk\backslash d}\right)}{\prod^{\mathcal{T}^{(t)}_{d}-1}_{u=0}\left[\sum^{N}_{n=1}\left(\alpha^{(t)}(n|k)+(\tau z)^{(t)}_{nk\backslash d}\right) + u \right]},
\end{align}
\end{widetext}
with
%% $Z^{(t)}_{k\backslash d} = \sum^{D}_{d'\neq d}z^{(t)}_{d'k}$
%% and
$(\tau z)^{(t)}_{nk\backslash d} = \sum^{D}_{d'\neq d}\tau^{(t)}_{nd'} z^{(t)}_{d'k}$
and
$\mathcal{T}^{(t)}_{d} = \sum^{N}_{n=1} \tau^{(t)}_{nd}$.
The derivation of the above equations is discussed in detail in App.~\ref{app: Collapsed Gibbs Sampling}.

In contrast to the Gibbs sampling algorithm, the probability distribution $p^{(t)}$, whose estimate determines the Dirichlet prior of $p^{(t+1)}$ at the next time step, is not directly sampled in the collapsed Gibbs sampling.
Instead, we can obtain the expectation value of $p^{(t)}(n|k)$ by using its posterior Dirichlet distribution as
\begin{align}
  \label{eq: p_t Collapsed Gibbs Sampling}
  p^{(t)}(n|k)
  &= \frac{\alpha^{(t)}(n|k) + \frac{1}{S}\sum^{S}_{s=1} \left(\tau z\right)^{(t)}_{nk;s}}{\sum^{N}_{n=1}\left[\alpha^{(t)}(n|k)+\frac{1}{S}\sum^{S}_{s=1}\left(\tau z\right)^{(t)}_{nk;s}\right]},
\end{align}
where $(\tau z)^{(t)}_{nk;s} = \sum^{D}_{d=1}\tau^{(t)}_{nd} z^{(t)}_{dk;s}$ is evaluated with realizations $z^{(t)}_{;s}$ of latent variables.
The update equations of $\alpha^{(t)}_{k}$ and $\eta^{(t)}_{k}$ are common to those in the Gibbs sampling.
In summary, pseudocode of the collapsed Gibbs sampling is shown in Algorithm~\ref{alg: Collapsed Gibbs Sampling}.

\begin{algorithm}[H]
  \caption{Collapsed Gibbs Sampling}         
  \label{alg: Collapsed Gibbs Sampling}
  \begin{algorithmic}[1]
    \STATE Initialize $p^{(0)}$, $\alpha^{(1)}$, and $\eta^{(1)}$
    \FOR{$t=1,2,\cdots,T_{\rm step}$}
    \STATE Initialize $z^{(t)}$
    \FOR{$s=1,2,\cdots,S$}
    \FOR{$d=1,2,\cdots,D$}
    \STATE Sample $z^{(t)}_{d}$ [Eq.~(\ref{eq: z Collapsed Gibbs Sampling})]
    \ENDFOR
    \ENDFOR
    \STATE Estimate $p^{(t)}$ [Eq.~(\ref{eq: p_t Collapsed Gibbs Sampling})]
    \STATE Compute $\alpha^{(t+1)}$ [Eq.~(\ref{eq: update of alpha})]
    \STATE Compute $\eta^{(t+1)}$ [Eq.~(\ref{eq: eta inference})]
    \ENDFOR
  \end{algorithmic}
\end{algorithm}

\subsubsection{Variational Bayesian Approach}

We can derive the lower bound of the log likelihood by using Jensen's inequality as
$\ln P(\tau^{(t)}| \alpha^{(t)},\eta^{(t)}) \geq F[q(z^{(t)},p^{(t)},\pi^{(t)})]$
with the variational lower bound
\begin{align}
  \label{eq: variational lower bound}
  &F[q(z^{(t)},p^{(t)},\pi^{(t)})]
  = \int dp^{(t)} d\pi^{(t)}  \sum_{z^{(t)}} 
  q(z^{(t)},p^{(t)},\pi^{(t)}) \nonumber \\
  & \hspace{40pt} \times
  \ln
  \frac{P(\tau^{(t)},z^{(t)},p^{(t)},\pi^{(t)} | \alpha^{(t)},\eta^{(t)})}{q(z^{(t)},p^{(t)},\pi^{(t)})},
\end{align}
which should be maximized with a distribution $q(z^{(t)},p^{(t)},\pi^{(t)})$.
We assume that an optimal form of the distribution can be decoupled as
\begin{align}
  \label{eq: mean-field approximation}
  q(z^{(t)},p^{(t)},\pi^{(t)})
  =
  q^{z}(z^{(t)}) q^{p}(p^{(t)}) q^{\pi}(\pi^{(t)}),
\end{align}
with
\begin{align}
  \label{eq: mean-field of q_z}
  & q^{z}(z^{(t)}) = \prod^{D}_{d=1} \prod^{K}_{k=1} q^{z}_{dk}(z^{(t)}_{dk}),\\
  \label{eq: mean-field of q_p}
  & q^{p}(p^{(t)}) = \prod^{K}_{k=1} q^{p}_{k}(p^{(t)}(:|k)),\\
  \label{eq: mean-field of q_pi}  
  & q^{\pi}(\pi^{(t)}) = \prod^{D}_{d=1} q^{\pi}_{d}(\pi^{(t)}_{d}).
\end{align}

With the aid of the joint probability distribution (\ref{eq: joint probability distribution}) and the mean-field approximation (\ref{eq: mean-field approximation}), the variational lower bound (\ref{eq: variational lower bound}) can be computed as
\begin{align}
  &F[q(z^{(t)},p^{(t)},\pi^{(t)})]
  \nonumber \\
  &= \int dp^{(t)} d\pi^{(t)} \sum_{z^{(t)}}
  q^{z}(z^{(t)}) q^{p}(p^{(t)}) q^{\pi}(\pi^{(t)}) \nonumber \\
  & \hspace{80pt}\times
  \ln P(\tau^{(t)}|z^{(t)},p^{(t)}) P(z^{(t)}|\pi^{(t)})
  \nonumber \\
  & \hspace{10pt}
  - D_{\rm KL}(q^{p}||P(p^{(t)}|\alpha^{(t)}))
  - D_{\rm KL}(q^{\pi}||P(\pi^{(t)}|\eta^{(t)}))  \nonumber \\
  & \hspace{10pt}
  - \sum_{z^{(t)}} q^{z}(z^{(t)}) \ln q^{z}(z^{(t)}),
\end{align}
with the Kullback-Leibler divergence
\begin{align}
  D_{\rm KL}(Q||P) = \int dx Q(x) \ln \frac{Q(x)}{P(x)}.
\end{align}

The optimal functional form of $q^{z}$, $q^{p}$, and $q^{\pi}$ can be obtained by functionally differentiating the variational lower bound (\ref{eq: variational lower bound}) with respect to them as
\begin{align}
  \label{eq: q z}
  & q^{z}_{dk}(z^{(t)}_{dk}=1)
  \propto
  \frac{\exp\left[\sum^{N}_{n=1}\tau^{(t)}_{nd} \Psi(\xi^{(t)}_{pnk}) \right]}{\exp\left[\sum^{N}_{n=1}\tau^{(t)}_{nd} \Psi(\sum^{N}_{n'=1}\xi^{(t)}_{pn'k}) \right]}
  \nonumber \\
  & \hspace{80pt} \times
  \frac{\exp\left[\Psi(\xi^{(t)}_{\pi dk})\right]}{\exp\left[\Psi(\sum^{K}_{k'=1}\xi^{(t)}_{\pi dk'})\right]},
  \\
  \label{eq: q pi}
  & q^{\pi}_{d}(\pi^{(t)})
  = {\rm Dir} (\pi^{(t)}_{d}| \xi^{(t)}_{\pi}),  \\
  \label{eq: q p}
  & q^{p}_{k}(p^{(t)}(:|k))
  = {\rm Dir} (p^{(t)}(:|k)| \xi^{(t)}_{p:k}),
\end{align}
where $\Psi(x)$ denotes the digamma function and the parameters are defined by
\begin{align}
  \label{eq: xi pi}
  &\xi^{(t)}_{\pi dk} = \eta^{(t)}_{k} + \gamma^{(t)}_{dk}, \\
  \label{eq: xi p}
  &\xi^{(t)}_{pnk}   = \alpha^{(t)}(n|k) +\sum^{D}_{d=1} \tau^{(t)}_{nd} \gamma^{(t)}_{dk},
\end{align}
with the responsibility
$\gamma^{(t)}_{dk} = q^{z}_{dk}(z^{(t)}_{dk}=1)$.
% The detailed derivation of Eqs.~(\ref{eq: q z})-(\ref{eq: xi p}) is given in App.~\ref{app: Variational Bayes}.
The parameters $\gamma^{(t)}_{dk}$, $\xi^{(t)}_{\pi dk}$, and $\xi^{(t)}_{pnk}$ in the variational Bayesian approach are self-consistently determined by solving the coupled equations (\ref{eq: q z}), (\ref{eq: xi pi}), and (\ref{eq: xi p}) at each time step.

Finally, we estimate $p^{(t)}$ by using its expectation value as
\begin{align}
  \label{eq: p_t VB estimate}
  p^{(t)}(n|k) 
  = \frac{\alpha^{(t)}(n|k) +\sum^{D}_{d=1} \tau^{(t)}_{nd} \gamma^{(t)}_{dk}}{\sum^{N}_{n=1}\left[ \alpha^{(t)}(n|k) +\sum^{D}_{d=1} \tau^{(t)}_{nd} \gamma^{(t)}_{dk} \right]},
\end{align}
at the end of each Markov step.
The optimization scheme based on the variational Bayesian approach is summarized in Algorithm~\ref{alg: Variational Bayes}.
In this paper, we repeat the substitution of the parameters for Eqs.~(\ref{eq: q z}), (\ref{eq: xi pi}), and (\ref{eq: xi p}) $T_{\rm iter}$ times on each Markov time step to achieve the self-consistent solutions.

\begin{algorithm}[H]  
  \caption{Variational Bayes}         
  \label{alg: Variational Bayes}
  \begin{algorithmic}[1]
    \STATE Initialize $\xi^{(1)}_{\pi}$, $\xi^{(1)}_{p}$, $p^{(0)}$, $\alpha^{(1)}$, and $\eta^{(1)}$
    \FOR{$t=1,2,\cdots,T_{\rm step}$}
    \FOR{$iter=1,2,\cdots,T_{\rm iter}$}
    \FOR{$d=1,2,\cdots,D$}
    \STATE Update $q^{z}_{dk}(z^{(t)}_{dk}=1)$ [Eq.~(\ref{eq: q z})]
    \STATE Update $\xi^{(t)}_{\pi d:}$ [Eq.~(\ref{eq: xi pi})]
    \ENDFOR
    \FOR{$k=1,2,\cdots,K$}
    \STATE Update $\xi^{(t)}_{p:k}$ [Eq.~(\ref{eq: xi p})]
    \ENDFOR
    \ENDFOR
    \STATE Estimate $p^{(t)}$ [Eq.~(\ref{eq: p_t VB estimate})]
    \STATE Compute $\alpha^{(t+1)}$ [Eq.~(\ref{eq: update of alpha})]
    \STATE Compute $\eta^{(t+1)}$ [Eq.~(\ref{eq: eta inference})]
    \ENDFOR
  \end{algorithmic}
\end{algorithm}

\subsubsection{EM Algorithm}
\label{subsec: EM Algorithm}

The original MDMC model developed by Okamoto and Qiu utilized the EM algorithm to infer the model parameters~\cite{mdmc_original,Okamoto2019ModularDO}.
In the following, we derive the EM algorithm based on the results of the variational Bayesian approach, and discuss differences between the original and current formulations.

We will first determine the probability $p^{(t)}(:|k)$ with the MAP estimation.
Since $p^{(t)}(:|k)$ obeys the Dirichlet distribution (\ref{eq: q p}), the MAP estimate of the parameter is given by the mode of the posterior Dirichlet distribution as
\begin{align}
  \label{eq: p_t point esimate}
  &p^{(t)}(n|k) 
  = \frac{1}{N_k}
  \left[
    \alpha^{(t)}_{k} \sum^{N}_{m=1} T_{nm}p^{(t-1)}(m|k)+\sum^{D}_{d=1} \tau^{(t)}_{nd} \gamma^{(t)}_{dk}
    \right].
\end{align}
Here, we have used the definition of $\alpha^{(t)}(n|k)$ [Eq.~(\ref{eq: definition of alpha})] and introduced the normalization factor
$N_{k} = \sum^{K}_{k=1} \left[\alpha^{(t)}_{k} \sum^{N}_{m=1} T_{nm}p^{(t-1)}(m|k)+\sum^{D}_{d=1} \tau^{(t)}_{nd} \gamma^{(t)}_{dk} \right]$.
At this stage, the designing principles of the prior parameter $\alpha^{(t)}(n|k)$ can be clarified.
The first term in the right hand side of Eq.~(\ref{eq: p_t point esimate}) corresponds to the master equation with the original network structure, while the second one describes the modification via $D$ times observation.
Hence, Eq.~(\ref{eq: p_t point esimate}) describes the Markovian dynamics of random walkers on a network augmented by the explicitly observed data $\tau^{(t)}$.
The hyperparameter $\alpha^{(t)}_{k}$ controls the ratio of these contributions.
When we take the infinite limit of the parameter as $\alpha^{(t)}_{k} \rightarrow \infty$, Eq.~(\ref{eq: p_t point esimate}) is reduced to the original master equation, where random walkers are independently obeying the same equation for each community.
On the other hand, the contributions from the random walkers are washed away in the vanishing limit of $\alpha^{(t)}_{k}$.

The current modeling of MDMC is slightly different from the original one~\cite{mdmc_original,Okamoto2019ModularDO} in the presence of the prior distribution of $\pi^{(t)}_{dk}$.
Since the variance of the Dirichlet distribution shrinks in large parameter regimes, the original model can be recovered by taking the infinite $\eta$ limit of ours.
Noting that the digamma function is asymptotically identical to the logarithm function as $\Psi(x) \sim \ln(x)$ for $x \gg 1$,
we can transform the result~(\ref{eq: q z}) of the variational Bayesian approach to
\begin{align}
  \label{eq: q z point-estimate}
  & \gamma^{(t)}_{dk}
  =
  \frac{\eta^{(t)}_{k} \text{Mult}\left( \tau^{(t)}_{:d}|p^{(t)}(:|k)\right)}{\sum^{K}_{k=1}\eta^{(t)}_{k} \text{Mult}\left( \tau^{(t)}_{:d}|p^{(t)}(:|k)\right)}.
\end{align}

The estimate of the probability distribution (\ref{eq: p_t point esimate}), the responsibility (\ref{eq: q z point-estimate}), and the update equation (\ref{eq: eta inference}) of $\eta^{(t)}_{k}$ can form a closed set of self-consistent equations, which are equivalent to those proposed in the original MDMC~\cite{mdmc_original,Okamoto2019ModularDO}.
We note that there remains some differences in the policies of updating the parameters $\alpha^{(t)}_{k}$ and $\eta^{(t)}_{k}$:
(1) The parameters $\alpha^{(t)}_{k}$ are not updated in the original MDMC.
(2) The parameters $\eta^{(t)}_{k}$ are updated at the end of each {\it{time step}} in the current formulation, while they are updated at the end of each {\it{EM step}} in the original MDMC.
In the following analysis, we use our new procedures described in Algorithm~\ref{alg: EM Algorithm} in order to compare the algorithm with other solvers.
%, in the same calculation condition.

\begin{algorithm}[H]  
  \caption{EM Algorithm}         
  \label{alg: EM Algorithm}
  \begin{algorithmic}[1]
    \STATE Initialize $p^{(0)}$, $\alpha^{(1)}$, and $\eta^{(1)}$
    \FOR{$t=1,2,\cdots,T_{\rm step}$}
    \FOR{$iter=1,2,\cdots,T_{\rm iter}$}
    \FOR{$d=1,2,\cdots,D$}
    \STATE Update responsibility $\gamma^{(t)}_{d}$ [Eq.~(\ref{eq: q z point-estimate})]
    \ENDFOR
    \FOR{$k=1,2,\cdots,K$}
    \STATE Update $p^{(t)}(:|k)$ [Eq.~(\ref{eq: p_t point esimate})]
    \ENDFOR
    \ENDFOR
    \STATE Estimate $p^{(t)}$ [Eq.~(\ref{eq: p_t point esimate})]
    \STATE Compute $\alpha^{(t+1)}$ [Eq.~(\ref{eq: update of alpha})]
    \STATE Compute $\eta^{(t+1)}$ [Eq.~(\ref{eq: eta inference})]
    \ENDFOR
  \end{algorithmic}
\end{algorithm}

    %% \FOR{$t=1,2,\cdots,T$}
    %% \STATE Initialize $z^{(t)}$
    %% \FOR{$s=1,2,\cdots,S$}
    %% \FOR{$d=1,2,\cdots,D$}
    %% \STATE Sample $z^{(t)}_{d}$ from the distribution (\ref{eq: z Collapsed Gibbs Sampling})
    %% \ENDFOR
    %% \ENDFOR
    %% \ENDFOR

\section{Evaluation of parameter inference methods}
\label{sec: evaluation}

\begin{figure*}[htbp]
 \begin{minipage}{0.47\hsize}
   \centering
   \subfigure[$o_{n}/N=0$]{
     \includegraphics[width=90mm]{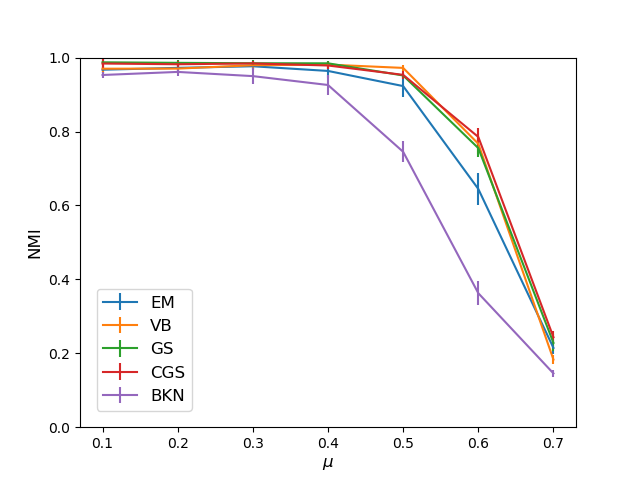}
     \label{fig: nmi on 0}}
 \end{minipage}
 \begin{minipage}{0.47\hsize}
   \centering
   \subfigure[$o_{n}/N=0.1$]{
     \includegraphics[width=90mm]{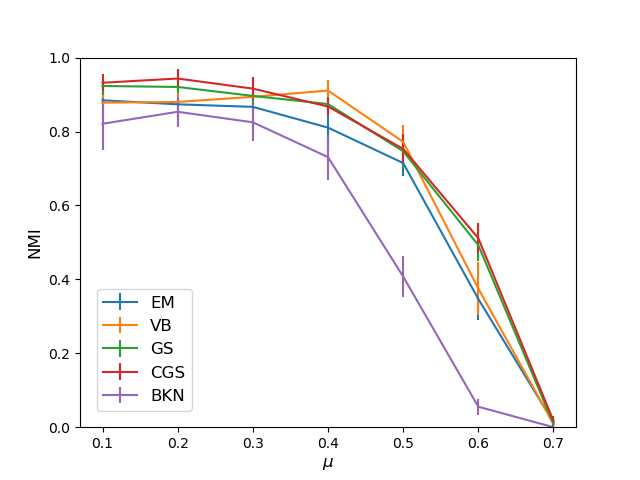}
     \label{fig: nmi on 0.1}}
 \end{minipage}
 \begin{minipage}{0.47\hsize}
   \centering
   \subfigure[$o_{n}/N=0.2$]{
     \includegraphics[width=90mm]{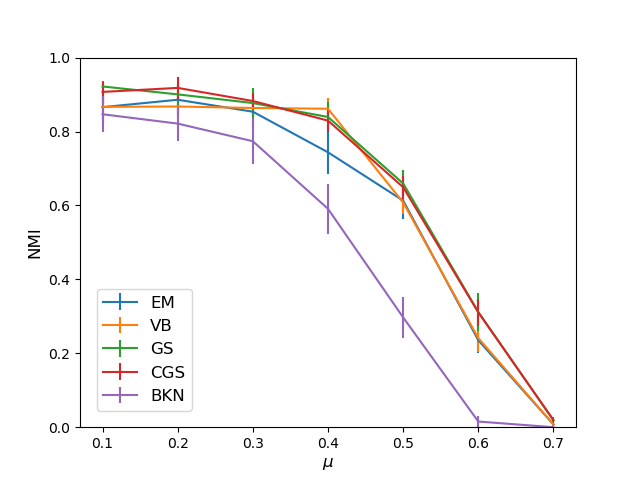}
     \label{fig: nmi on 0.2}}
 \end{minipage}
 \begin{minipage}{0.47\hsize}
   \centering
   \subfigure[$o_{n}/N=0.3$]{
     \includegraphics[width=90mm]{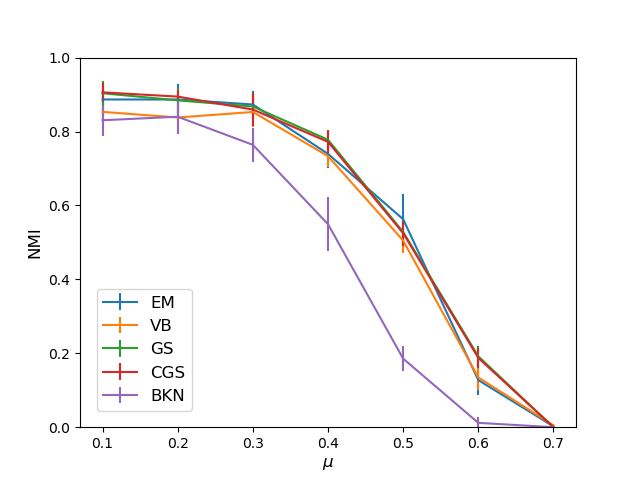}
     \label{fig: nmi on 0.3}}
 \end{minipage}
 \caption{
   (Color online)
   The results of the extended NMI for LFR benchmark with various values of the mixing parameter $\mu$ and the number $o_{n}$ of the overlapping nodes.
   The networks are generated $10$ times for each parameter set with
   $N=1000$, $o_{m}=2$, $t_{1}=2$, $t_{2}=1$,
   $k_{\rm ave}=15$, $k_{\rm max}=50$, $c_{\rm min}=20$, and $c_{\rm max}=100$.
   MDMC is performed with
   $T_{\rm step}=50$, $\alpha^{(1)}_{k=1,2,\cdots,K}/D=0.1$, and $\eta^{(1)}_{k=1,2,\cdots,K}=1$,
   where $D$ and $K$ are read from each generated network.
   The number of the Monte Carlo sampling is set as $S=100$ with the burn in period $S_{\rm burn}=100$ in the Gibbs samplers,
   while the number of the iteration is set as $T_{\rm iter}=1000$ and $100$ for the EM algorithm and the variational Bayesian approach, respectively.
 }
 \label{fig: LFR nmi}
\end{figure*}

In this section, we apply the optimization algorithms developed in the previous section to various testbed networks to qualitatively evaluate their performance.
We use a normalized mutual information (NMI), which measures the similarity between two different partitions of community assignments, namely our prediction and the ground truth.
As has been discussed in the previous section, modular decomposition of Markov chain (MDMC) has been designed to capture global structures of networks by combining stochastic modeling of network generation processes and Markovian dynamics of random walkers.
%% In the context of Markovian dynamics, the community-detection results at each time slice can be considered as Markovian dynamics of the random walker.
%% This interpretation allows us to visualize and diagnose the optimization processes of the parameter inference by tracking the probability distribution of the agent.
This interpretation allows us to further analyze the detected communities by visualizing and diagnosing optimization processes.

% This idea is similar to the PageRank algorithm [citation], where the centrality of nodes are evaluated by solving the eigenvalue problem of a stochastic matrix representing the world-wide webs.

% in order to further analyze the results obtained by MDMC.

\subsection{Node Clustering}

While MDMC assigns communities to every {\it link} in networks, most of the existing testbed networks are equipped with ground-truth {\it node} communities.
Hence, we need an additional policy to determine the membership of the nodes within our framework.
In this paper, we will define the joint probability distribution of the random walker with respect to node $n$ and community $k$ as
$p^{(t)}(n,k) = p^{(t)}(n|k)p^{(t)}(k)$ with $p^{(t)}(k)=\eta^{(t)}_{k}/\sum^{K}_{k=1}\eta^{(t)}_{k}$.
Then, the most probable community can be assigned to each node $n$ by taking the largest component of $p^{(t)}(k|n)$.
As a metric for comparison, we use an NMI which has been extended to quantify similarities of two different partitions with overlapping communities~\cite{Lancichinetti_2009,McDaidNMI}.

\subsection{Experiment with LFR Benchmark}
\label{subsec: experiment with LFR benchmark}

We use LFR benchmark~\cite{PhysRevE.78.046110} in order to evaluate the performance of MDMC solvers.
The LFR benchmark has been widely used to evaluate community detection algorithms because it allows us to generate various types of realistic networks including those with overlapping communities~\cite{PhysRevE.80.056117,Xie:2013:OCD:2501654.2501657,ding_overlapping_2016}.
We will compare MDMC with a stochastic model, which was proposed by Ball {\it et al.} in Ref.~\onlinecite{ball2011efficient} and has been considered as one of the best models to detect overlapping communities.
% denoted by "BKN" model hereafter.

%% generation of LFR benchmark
We have generated undirected and unweighted networks with $N=1000$ nodes
for four different ratios [$o_{n}/N=0$, $0.1$, $0.2$, and $0.3$] of nodes on which $o_{m}=2$ different communities overlap.
The minus exponents of the degree and the community size distributions are set as $t_{1}=2$ and $t_{2}=1$, respectively.
The average degree is $k_{\rm ave}=15$ with an upper bound $k_{\rm max}=50$.
The size of the generated communities ranges from $c_{\rm min}=20$ to $c_{\rm max}=100$.
For each overlapping-node ratio, the mixing parameter $\mu$, which controls the number of edges between different communities, is varied from $\mu=0.1$ to $0.7$ with spacing $0.1$ to examine effects of the mixtures on the performance.
%% experimental setup
We generated $10$ networks for each parameter set to evaluate the mean and the variance of NMI values.
%% parameters in MDMC
In order to set initial values of MDMC parameters, we read the number of ground truth community $K$ and that of edges $D$ for each network and set the values as $\alpha^{(1)}_{k}/D= 0.1$ and $\eta^{(1)}_{k}=1$ for $k=1,2,\cdots,K$.

We show the results of the extended NMI for LFR benchmark in Fig.~\ref{fig: LFR nmi}, where EM, VB, GS, and CGS stand for the EM algorithm, the variational Bayesian approach, the Gibbs sampling, and the collapsed Gibbs sampling, respectively.
In wide parameter regimes, the Gibbs samplers outperform the variational Bayesian approach and EM algorithms in terms of the NMI values.
In particular, the results of the deterministic methods deteriorate as the number of overlapping nodes increases, while the Gibbs samplers keep their qualities ($0.8 \lesssim$ NMI) even in highly overlapping $o_{n}/N=0.3$ and mixed $\mu \lesssim 0.4$ situations.
The Gibbs and collapsed Gibbs sampling algorithms give the same values within statistical errors in every situation, which indicates that the number $S$ of Monte Carlo sampling and the additional parameter $S_{\rm burn}$ denoting the burn-in period are large enough to achieve the best performance of the model.
The results of Ball's model~\cite{ball2011efficient}, which is denoted by BKN, is comparable or slightly inferior to those from MDMC with EM-algorithm solver.
There may be two possible reasons to explain this behavior: representability of these models and performance of their solvers.
MDMC is designed to incorporate global structures of networks by introducing random walkers, while Ball's model essentially focuses on local link structures around each node.
%% Hence, our  MDMC model may capture important network characteristics, which are absent in Ball's model.
Another possible reason is that Ball's model exploits a simple EM algorithm, which is known to suffer from trapping of locally optimized states.
This leads to a natural concern that the EM algorithm might have not utilized the maximum potential of Ball's model.
In order to fully answer this question, it is necessary to systematically elaborate Ball's model, which is out of scope of this paper but deserves another research.

\subsection{Real-world Network}

\begin{figure*}[htbp]
 \begin{minipage}{0.3\hsize}
   \centering
   \subfigure[$t=1$]{
     \includegraphics[width=50mm]{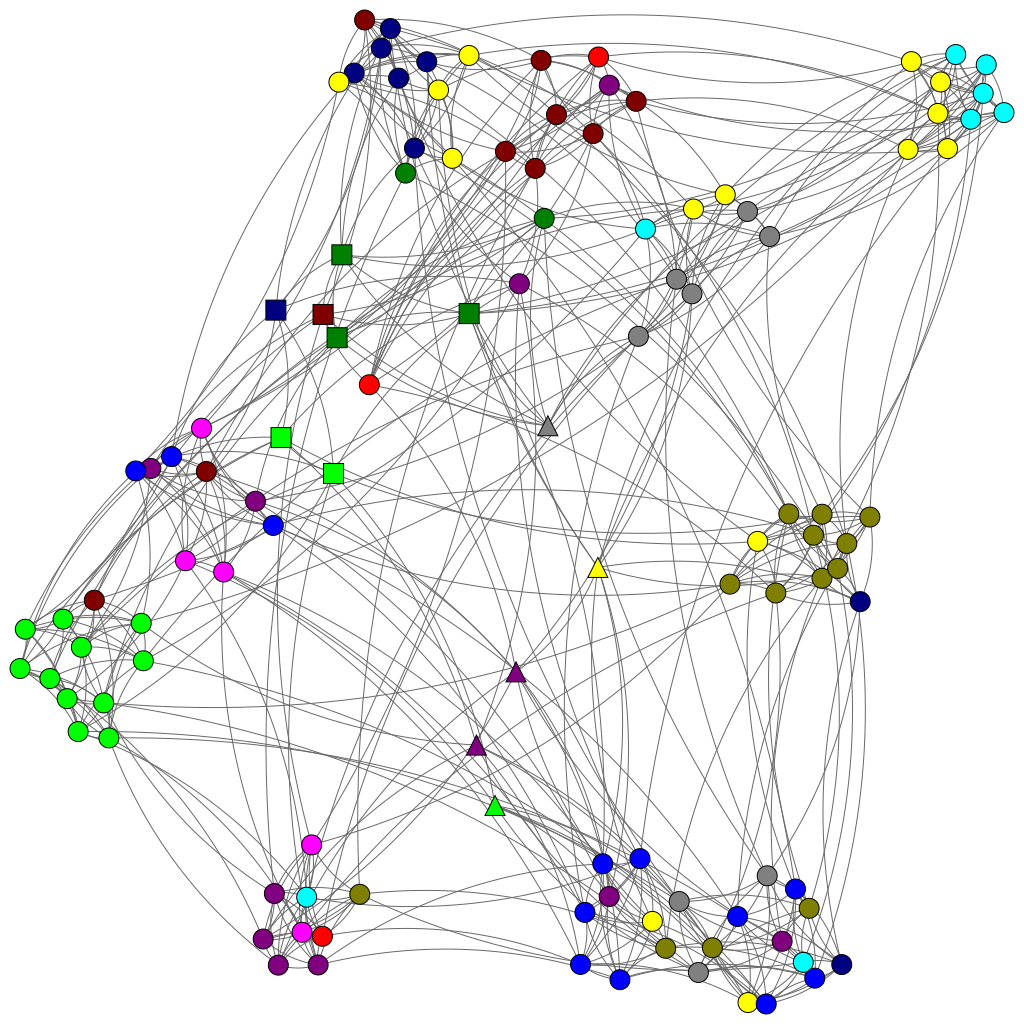}
     \label{fig: football t1}}
 \end{minipage}
 \begin{minipage}{0.3\hsize}
   \centering
   \subfigure[$t=3$]{
     \includegraphics[width=50mm]{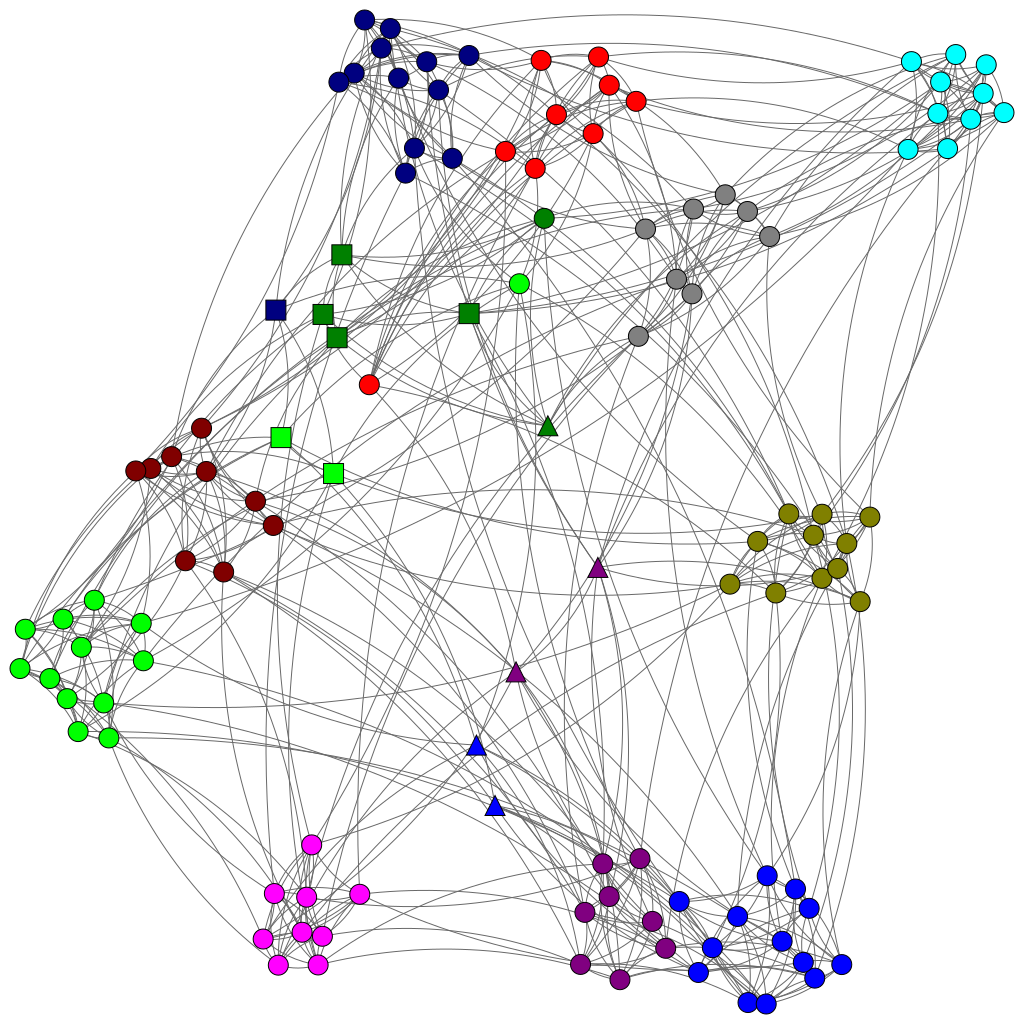}
     \label{fig: football t3}}
 \end{minipage}
 \begin{minipage}{0.3\hsize}
   \centering
   \subfigure[$t=50$]{
     \includegraphics[width=50mm]{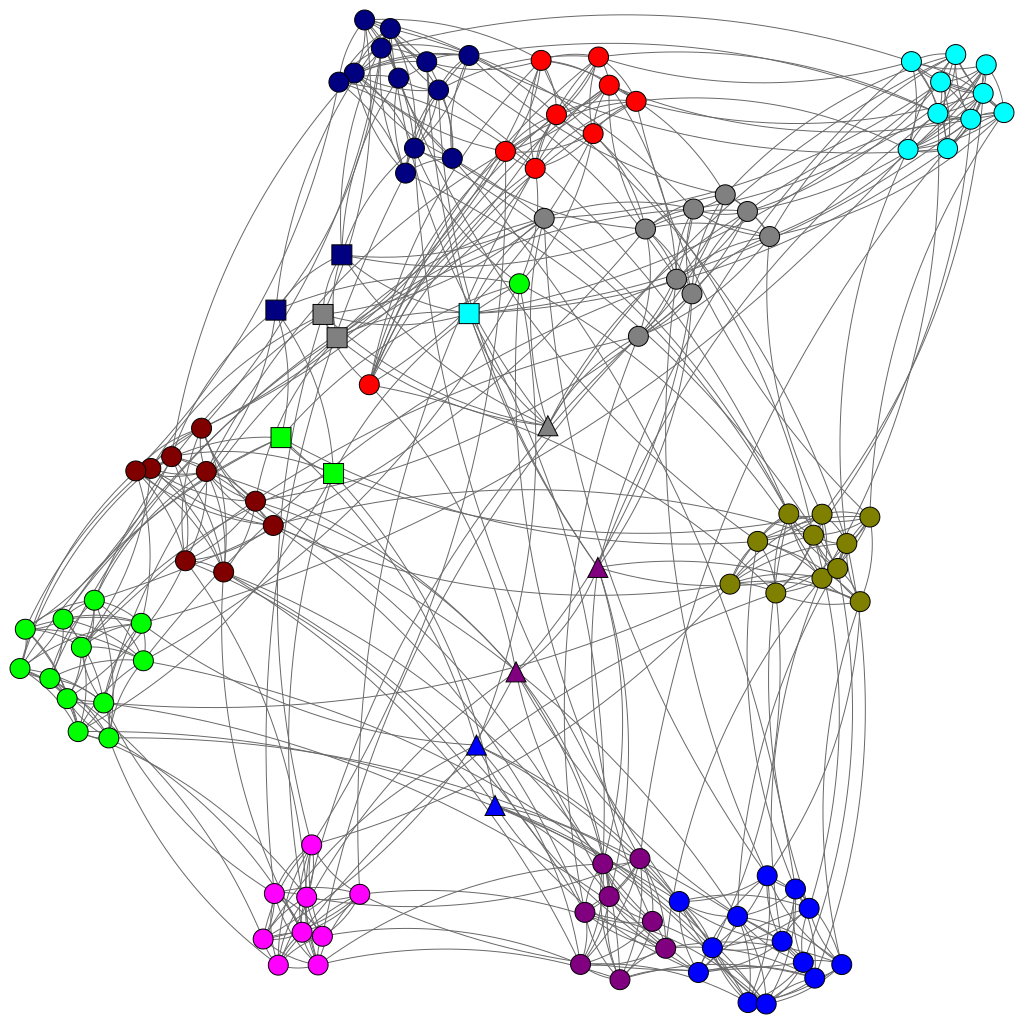}
     \label{fig: football t50}}
 \end{minipage}
 \caption{
   (Color online)
   Typical results of community assignments in American college football network at different time slices $t=1$, $3$, and $50$.
   The colors of the nodes represent their main communities.
   The triangle and square nodes represent teams in ``Independents" and ``Sun Belt", respectively.
   MDMC is performed with 
   $T_{\rm step}=50$, $\alpha^{(1)}_{k=1,2,\cdots,K}/D=0.1$, $\eta^{(1)}_{k=1,2,\cdots,K}=1$, and $K=12$.
   The collapsed Gibbs sampling is used as an MDMC solver with $S_{\rm burn}=200$ and $S=1000$.
 }
 \label{fig: football pk}
\end{figure*}

\begin{figure*}[htbp]
 \begin{minipage}{0.3\hsize}
   \centering
   \subfigure[$\alpha^{(t)}_{k}/D$]{
     \includegraphics[width=60mm]{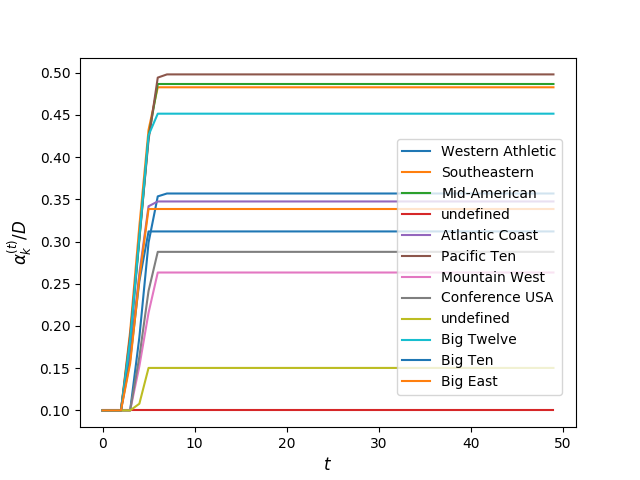}
     \label{fig: football alpha}}
 \end{minipage}
 \begin{minipage}{0.3\hsize}
   \centering
   \subfigure[$\eta^{(t)}_{k}$]{
     \includegraphics[width=60mm]{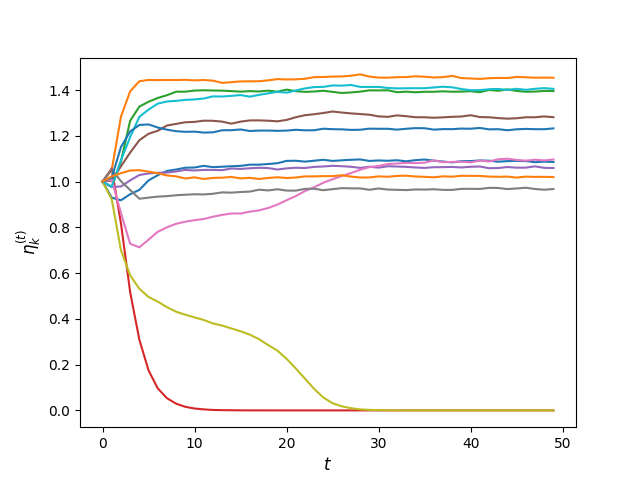}
     \label{fig: football eta}}
 \end{minipage}
 \begin{minipage}{0.3\hsize}
   \centering
   \subfigure[$\ln P$]{
     \includegraphics[width=60mm]{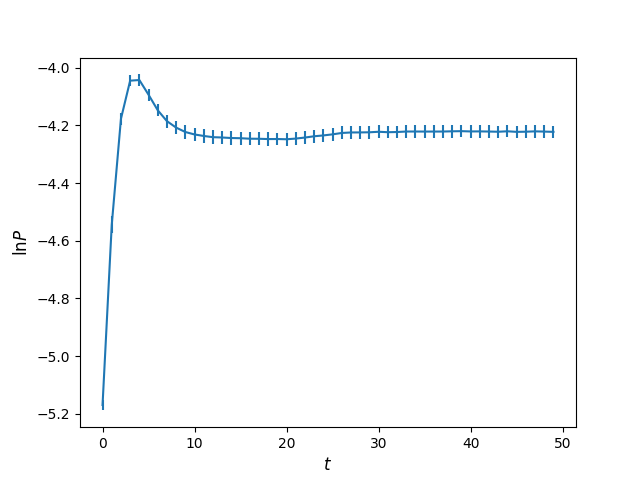}
     \label{fig: football logP}}
 \end{minipage}
 \caption{
   (Color online)
   The optimization processes of the hyperparameters (a) $\alpha^{(t)}_{k}$ and (b) $\eta^{(t)}_{k}$ and (c) the log likelihood $\ln P$ for American college football network with the same run used in Fig.~\ref{fig: football pk}.
   The labels of the colored curves are determined with the Jaccard index, and are common in (a) and (b).
 }
 \label{fig: football parameter}
\end{figure*}

It is meaningful to apply and examine our methods with real-world networks because they may have features which are not captured in synthetic ones.
Among the widely studied real-world networks, we choose American college football network~\cite{Girvan7821} to test the performance of our model.
The nodes and links in the dataset represent football teams and games played by two of them, respectively.
Every node in the network belongs to one of 12 communities called ``conferences", and games took place more often between teams in the same conference than those in different conferences.
Football teams in ``Independents" conference are exceptional because they do not have particular preferences for their opponents.
Since the generative processes of the network imply typical characteristics of node communities, link clustering schemes are known to suffer from severe challenges for detecting the ground truth communities~\cite{He2015}.
This observation raises a concern that MDMC, which models community assignments of links rather than nodes, may not detect the true communities.

\begin{table}[b]
  \caption{
    \label{tab: football}
    The statistics of NMI values and the counts of the number of detected main communities out of $100$ trials for American college football network.
    MDMC is performed with
    $T_{\rm step}=50$, $\alpha^{(1)}_{k}/D=0.1$, $\eta^{(1)}_{k}=1$, $D=613$, and $K=12$.
    The number of the Monte Carlo sampling and the burn in period is set as $S=1000$ and $S_{\rm burn}=200$, respectively in the Gibbs samplers.
    The number of the iteration is set as $T_{\rm iter}=1000$ ($100$) for the EM algorithm (variational Bayesian approach).
  }
  \begin{ruledtabular}
    \begin{tabular}{l||lrrrrr}
      \textrm{Method}&
      \textrm{NMI}&
      \textrm{$K_{\rm fin}=9$}&
      \textrm{$10$}&
      \textrm{$11$}&
      \textrm{$12$}\\
      \colrule
      EM  & 0.869 (0.018) & 37 & 53 & 10 & 0\\
      VB  & 0.871 (0.016) & 37 & 63 &  0 & 0\\
      GS  & 0.875 (0.004) &  0 & 92 &  8 & 0\\
      CGS & 0.875 (0.003) &  0 & 96 &  4 & 0\\
      BKN & 0.897 (0.021) &  0 &  0 & 20 & 80\\
    \end{tabular}
  \end{ruledtabular}
\end{table}

We have applied MDMC solvers to American college football, and summarized the mean (standard deviation) of the extended NMI in the second column in Table~\ref{tab: football}.
We have also analyzed the number of communities which are detected as main one of at least one node in the network at $t=T_{\rm step}=50$.
The counts of the final results with $K_{\rm fin}$ main communities out of $100$ trials are shown in the last four columns in Table~\ref{tab: football}.
The NMI scores from MDMC solvers are by a wide margin larger than the value 0.8035 reported by a previous method for node-link communities \cite{He2015},
while they are slightly smaller than the value obtained by Ball's model.
In terms of the standard deviation, the results of the deterministic approaches, i.e. the EM algorithm, the variational Bayesian approach, and Ball's model, are more fluctuating than those from the Gibbs samplers because they are subject to random initial states.
Crucial differences between these approaches appear in the number $K_{\rm fin}$ of main communities.
In particular, the number of the ground-truth communities $K=12$ is not reproduced in MDMC, while Ball's model predicted the results with $K_{\rm fin}=12$ main communities $80$ times out of $100$ trials.

In order to interpret the result that the number of main communities detected by MDMC is less than that of the ground truth communities, we identify the communities missed by our detection algorithms.
We establish the correspondence between the detected and the ground-truth communities by associating each detected community with the most similar ground-truth community in terms of Jaccard index.
The detected community is considered as ``undefined" when the maximum value of the Jaccard index is less than $0.5$.
With these criteria, we have found that ``Independents" and ``Sun Belt" communities are missing in most cases.
In later analysis, we use the collapsed Gibbs sampling and detail the missing communities by directly visualizing optimization processes of the community assignments.

Fig.~\ref{fig: football pk} shows typical results of the main community assignment $\mathop{\rm argmax}\limits_{k} p^{(t)}(k|n)$ at different time slices $t=1$, $3$, and $50$.
The colors of the nodes represent the main communities detected by our model.
Football teams belonging to ``Independents" and ``Sun Belt" conferences are shown as triangle and square nodes, respectively.
Since ``Independents" conference do not have particular community structures, the automatic drop of this community during optimization processes is favorable property of our model.
On the other hand, ``Sun Belt" nodes partially found at the early stage of the optimization [Fig.~\ref{fig: football t3}] looks washed away during the Markov chain processes [Fig.~\ref{fig: football t50}].
Considering that ``Sun Belt" is the smallest community aside from ``Independents", we infer that the resolution-limit problem, which are recognized in modularity optimization methods~\cite{fortunato2007resolution}, also matters in the MDMC approach.
In fact, small communities observed in early times of the optimization are prone to be merged into larger ones as the random walkers travel the network.

Further insights on the selection of optimal communities can be gained from optimization processes of the model parameters.
The left, center, and right panels of Fig.~\ref{fig: football parameter} show values of the hyperparameters $\alpha^{(t)}_{k}/D$ and $\eta^{(t)}_{k}$, and the log likelihood $\ln P^{(t)}$, respectively, obtained in the same run for Fig.~\ref{fig: football pk}.
The labels of the colored curves are assigned with the Jaccard index computed with communities finally detected at $t=50$, and are common in Figs.~\ref{fig: football alpha} and \ref{fig: football eta}.
As is consistent with the optimization of community assignments [Figs.~\ref{fig: football t1} and \ref{fig: football t3}], the parameters $\alpha^{(t)}_{k}$ and $\eta^{(t)}_{k}$ for most of the distinct communities are rapidly relaxed to the stationary values.
These fast optimization processes take place in just a few Markov steps ($t \lesssim 5$) because random walkers can diffuse within each community in the time-scale comparable to its average path length.
Subsequent slow dynamics occurs when the random walkers travel across different communities.
This induces global organization of communities as is observed in Fig.~\ref{fig: football eta} for $5 \lesssim t \lesssim 30$.
The parameters are finally converged into the stationary values in accordance with the basic principle of Markov chains~\footnote{We note that the convergence of the distributions is not guaranteed in general because of the stochastic optimization of the parameters.}.
These pictures also explain the behavior of the log likelihood in Fig.~\ref{fig: football logP}, where the error bars represent the mean and the standard deviation of $\ln P$ estimated by Monte Carlo sampling for each time step.
We note that the log likelihood is not monotonically increasing in time because of the approximation described in Sec.~\ref{subsec: approximation of likelihood function}.
Developing efficient algorithms to optimize the whole Markov chain of MDMC is one of the important remaining issues.

% meta-stable state in our model.
% Markov chain vs stochastic modeling.

%the state with small communities are meta-stable in a network sense.
%% Since the main belongings at early times are significantly dependent on the initial conditions, we consider that ``Sun Belt" community is meta-stable in a network sense, and merged into

\section{Discussions and Conclusions}
\label{sec: conclusions}

%% Community detection algorithms are essencial in network science because it eludicates .
%% Modular decomposition of Markov Chain (MDMC) proposed by Okamoto and Qiu~\onlinecite{mdmc_original} is one of the community detection methods.

In this paper, we have improved modular decomposition of Markov chain (MDMC), which is a stochastic model proposed in Ref.~\onlinecite{mdmc_original,Okamoto2019ModularDO} for detecting community structures in networks, by developing various optimization algorithms based on variational Bayesian and Monte Carlo sampling approaches.
We have applied MDMC to LFR benchmark~\cite{PhysRevE.78.046110}, and found that the Gibbs sampling algorithms outperform the EM algorithm, which was used in the original paper~\cite{mdmc_original,Okamoto2019ModularDO}, in wide parameter regimes.
In particular, overlapping communities are more accurately detected by the Gibbs samplers, which enhance the effectiveness of MDMC in elucidating these challenging community structures.

We have also examined the performance of MDMC with American college football network~\cite{Girvan7821}.
Real-world networks often possess properties which are absent in synthetic ones.
For instance, ground-truth communities are not necessarily consistent with given network structures as is the case with ``Independents" conference in American college football network.
We have found that ``Independents" community is automatically dropped during optimization processes of MDMC, while another promising stochastic model developed by Ball {\it et al.}~\cite{ball2011efficient} fails to exclude this irrelevant community.
The optimal community structures were more often detected by the Gibbs samplers, while deterministic methods, i.e. EM algorithm and the variational Bayesian approach, are too sensitive to initial conditions to robustly find them.
We have further detailed the optimization processes of MDMC parameters, and clarified that the automatic selection of the optimal communities is a consequence of global dynamics of random walkers on the network.

Although the Gibbs samplers developed in this paper allow us to exploit the full potential of MDMC to obtain accurate and stable results, they also clarify some limitations of the model.
First, the resolution-limit problem looks inevitable in the current formulation because small communities tend to be aggregated into larger ones as the random walkers travel through a network.
In order to detect locally-stable but globally-unstable community structures, we need to develop some schemes to reliably extract meta-stable structures in the optimization processes.
Second, the likelihood function is not guaranteed to monotonically increase in the Markov steps because it is separately optimized at different time slices [see Sec.~\ref{subsec: approximation of likelihood function} for detail of the approximation].
As a consequence, it is impossible at the present stage to simultaneously satisfy the two guiding principles of MDMC, i.e. the maximization of the likelihood function and the convergence to the steady-state distribution in the Markov chain.
It seems that reliable results are already achieved in a practical point of view, but efficient optimization methods unifying these guiding principles are desirable to assess the validity of this approximation.

It is shown, in this paper, that stochastic modeling of network generation processes are quite efficient to detect community structures in a wide variety of networks.
This is reminiscent of the situation in topic modeling, where latent Dirichlet allocation (LDA) model~\cite{blei2003latent} has ignited subsequent works on document clustering methods with additional structures~\cite{Blei:2006:DTM:1143844.1143859,Iwata:2009:TTM:1661445.1661674}.
Among the fundamental characteristics of real-world networks, the existence of underlying hierarchical structures is essential in many cases to understand intrinsic nature of the networks~\cite{ravasz2003hierarchical}.
Another promising direction is to utilize additional meta-information of nodes and links for community detection.
%%% idea?
We leave these promising directions for future works.

%% One of the most important properties of the MDMC is its extensibility to incorporate various aspects of the network by modeling the underlying structure of the networks.

%% The Bayesian formulation allows us to extend the stochastic models.

%% temporal network.
%% semisupervised, hierarchy, community-size prediction.

\begin{acknowledgments}
  The author would like to thank Xule Qiu, Seiya Inagi, Hiroshi Okamoto, Hiroshi Umemoto, and Takeshi Onishi for useful discussions and comments.
  In order to evaluate the proposed methods, an open software available at \url{https://sites.google.com/site/santofortunato/inthepress2} was used to generate synthetic networks.
  In addition, an extended normalized mutual information was computed with another open software available at \url{https://github.com/aaronmcdaid/Overlapping-NMI}.
\end{acknowledgments}

\appendix

\section{Derivation of Gibbs Sampling}
\label{app: Gibbs Sampling}

The probability of the latent variable $z^{(t)}_{dk}$ can be computed as
\begin{align}
  &P(z^{(t)}_{dk}=1|\tau^{(t)},z^{(t)}_{\backslash d},p^{(t)},\pi^{(t)},\alpha^{(t)},\eta^{(t)})
  \nonumber \\
  &=P(z^{(t)}_{dk}=1|\tau^{(t)}_{:d},p^{(t)},\pi^{(t)}_{d}),
  \nonumber \\
  &= \frac{P(\tau^{(t)}_{:d},z^{(t)}_{dk}=1|p^{(t)},\pi^{(t)}_{d})}{P(\tau^{(t)}_{:d}|p^{(t)},\pi^{(t)}_{d})},
  \nonumber \\
  &\propto P(\tau^{(t)}_{:d},z^{(t)}_{dk}=1|p^{(t)},\pi^{(t)}_{d}),
  \nonumber \\
  &\propto P(\tau^{(t)}_{:d},|z^{(t)}_{dk}=1,p^{(t)})
  P(z^{(t)}_{dk}=1|\pi^{(t)}_{d}),
  \nonumber \\
  &\propto \text{Mult}\left( \tau^{(t)}_{:d}|p^{(t)}(:|k)\right) \pi^{(t)}_{dk} .
\end{align}
Here, we have ignored the denominator of the third line because it is independent of $z^{(t)}_{dk}$.
In the last line, we have used Eqs.~(\ref{eq: pz}) and (\ref{eq: tau}).
The equation above is equivalent to Eq.~(\ref{eq: latent variable probability}) up to a normalization factor.

The posterior probability of $p^{(t)}(:|k)$ is computed as
\begin{align}
  &P(p^{(t)}(:|k)|\tau^{(t)},z^{(t)},p^{(t)}(:|\backslash k),
  \pi^{(t)},\alpha^{(t)},\eta^{(t)}) \nonumber \\
  & \propto
  P(\tau^{(t)},p^{(t)}(:|k)|
  z^{(t)},\alpha^{(t)}), \nonumber \\
  &\propto
  \left[
    \prod^{D}_{d=1}
    \left( \text{Mult}\left( \tau^{(t)}_{:d}|p^{(t)}(:|k)\right) \right)^{z^{(t)}_{dk}}
    \right]
  \nonumber \\
  & \hspace{20pt} \times
  \text{Dir}(p^{(t)}(:|k)|\alpha^{(t)}(:|k)),
  \nonumber \\
  &\propto \prod^{N}_{n=1} \left[p^{(t)}(n|k)\right]^{\alpha^{(t)}(n|k) + (\tau z)^{(t)}_{nk}-1}.
\end{align}
In the third line, we have used Eqs.~(\ref{eq: pt}) and (\ref{eq: tau}), and left the factor relevant to $p^{(t)}(:|k)$ in the subsequent line.
Noting that the last line is proportional to the Dirichlet distribution with parameter $\alpha^{(t)}(n|k) + (\tau z)^{(t)}_{nk}$, we arrive at the expression (\ref{eq: p_t probability}).

As is seen from the graphical model [Fig.~\ref{fig: graphical representation}], the probability distribution of $\pi^{(t)}_{d}$ depends only on the parameter $\eta^{(t)}$ of the prior distribution and the latent variables $z^{(t)}_{d}$.
Since the Dirichlet distribution is a conjugate prior of the multinomial distribution, the resulting probability distribution is the Dirichlet distribution with a parameter $\eta^{(t)}+z^{(t)}_{d}$.
This is nothing but the probability distribution (\ref{eq: pi_t probability}).

\section{Derivation of Collapsed Gibbs Sampling}
\label{app: Collapsed Gibbs Sampling}

The marginalized joint probability distribution is rewritten as
\begin{align}
  & P(\tau^{(t)},z^{(t)}|\alpha^{(t)},\eta^{(t)})
  \nonumber \\
  \label{eq: decomposition of marginalized dist.}
  &  = 
  P(\tau^{(t)}|z^{(t)},\alpha^{(t)},\eta^{(t)})
  P(z^{(t)}|\alpha^{(t)},\eta^{(t)}),
\end{align}
with the multiplication rule.
In the following, we will separately compute the two factors in the last line of the above equation.

The first factor in Eq.~(\ref{eq: decomposition of marginalized dist.}) is computed as
\begin{align}
  &P(\tau^{(t)}|z^{(t)},\alpha^{(t)},\eta^{(t)}) \nonumber \\
  & =
  \prod^{K}_{k=1}
  \int dp^{(t)}(:|k) 
   \left( \prod^{D}_{d=1} \text{Mult}\left( \tau^{(t)}_{:d}|p^{(t)}(:|k)\right)^{z^{(t)}_{dk}}\right)
    \nonumber \\    
    &\hspace{90pt} \times 
    \text{Dir}(p^{(t)}(:|k)|\alpha^{(t)}(:|k))
  ,
  \nonumber \\
  \label{eq: first factor in marginalized dist.}
  & = \prod^{K}_{k=1} \frac{\Gamma\left( \sum^{N}_{n=1} \alpha^{(t)}(n|k) \right)}{\prod^{N}_{n=1} \Gamma(\alpha^{(t)}(n|k))}
  \nonumber \\    
  &\hspace{30pt} \times 
  \frac{\prod^{N}_{n=1}\Gamma\left( \alpha^{(t)}(n|k)+ (\tau z)^{(t)}_{nk}\right)}{\Gamma\left(\sum^{N}_{n=1} \left( \alpha^{(t)}(n|k)+ (\tau z)^{(t)}_{nk}\right)\right)}.
\end{align}
The last line of Eq.~(\ref{eq: first factor in marginalized dist.}) is obtained with the identity
\begin{equation}
  \int d\phi \prod^{N}_{n=1} \phi^{\beta_{n}-1}_{n}
  = \frac{\prod^{N}_{n=1} \Gamma \left(\beta_{n}\right)}
  {\Gamma \left(\sum^{N}_{n=1}\beta_{n}\right)},
\end{equation}
where $\int d\phi$ denotes the integral over the probability distribution $\phi_{n=1,2,\cdots,N}$.
The second factor in Eq.~(\ref{eq: decomposition of marginalized dist.}) is computed in a similar way as
\begin{align}
  &P(z^{(t)}|\alpha^{(t)},\eta^{(t)}) \nonumber \\
  & =
  \prod^{D}_{d=1}
  \int d\pi^{(t)}_{d}
  \left[
    \prod^{K}_{k=1}
     \left( \pi^{(t)}_{dk}\right)^{z^{(t)}_{dk}}
    \right]
    \text{Dir}\left(\pi^{(t)}_{d}|\eta^{(t)}\right)
  ,
  \nonumber \\
  \label{eq: second factor in marginalized dist.}
  & = 
  \prod^{D}_{d=1}
  \frac{\Gamma\left( \sum^{K}_{k=1} \eta^{(t)}_{k} \right)}{\prod^{K}_{k=1} \Gamma\left(\eta^{(t)}_{k}\right)}
  \frac{\prod^{K}_{k=1}\Gamma\left( \eta^{(t)}_{k}+ z^{(t)}_{dk}\right)}{\Gamma\left(\sum^{K}_{k=1} \left( \eta^{(t)}_{k}+ z^{(t)}_{dk}\right)\right)}.
\end{align}
Multiplying the first (\ref{eq: first factor in marginalized dist.}) and the second (\ref{eq: second factor in marginalized dist.}) factors, we have obtained the marginalized distribution (\ref{eq: Marginalized likelihood function CGS}) described in the main text.

The probability distribution of the latent variable $z^{(t)}$ is calculated as
\begin{align}
  & P(z^{(t)}_{dk}=1|\tau^{(t)},z^{(t)}_{\backslash d},\alpha^{(t)},\eta^{(t)})
  \nonumber \\
  &\propto P(\tau^{(t)}_{:d},z^{(t)}_{dk}=1|\tau^{(t)}_{:\backslash d},z^{(t)}_{\backslash d},\alpha^{(t)},\eta^{(t)}),
  \nonumber \\
  \label{eq: CGS Appendix 1}
  &\propto \frac{P(\tau^{(t)},z^{(t)}_{dk}=1,z^{(t)}_{\backslash d}|\alpha^{(t)},\eta^{(t)})}{P(\tau^{(t)}_{:\backslash d},z^{(t)}_{\backslash d}|\alpha^{(t)},\eta^{(t)})},
\end{align}
up to a constant factor in terms of $z^{(t)}$.
Most of the contributions in the numerator and denominator of Eq.~(\ref{eq: CGS Appendix 1}) are canceled out because they are nothing but the marginalized likelihood function (\ref{eq: Marginalized likelihood function CGS}) with and without the observed data $\tau^{(t)}_{:d}$, respectively.
With the identity $\Gamma(x+1) = x \Gamma(x)$ and $\tau^{(t)}_{nd} \in \{0,1\}$, we have proved the distribution (\ref{eq: z Collapsed Gibbs Sampling}) for the collapsed Gibbs sampling.

By differentiating the explicit expression of the likelihood function (\ref{eq: Marginalized likelihood function CGS}), we can derive the update equation (\ref{eq: F_alpha}) for the parameter $\alpha^{(t)}$ with the relation of the of the digamma function $\Psi(x)$;
$\Psi(x+n)-\Psi(x)=\sum^{n}_{l=1}\frac{1}{x+l-1}$ for $n \in Z^{+}$.

%% \section{Derivation of Variational Bayes}
%% \label{app: Variational Bayes}

%% \section{Useful inequalities for Gamma distribution}
%% \label{app: Inequalities for Gamma Distribution}

%% For any $\hat{x}\geq 0$, $x>0$, $n \geq 0$,
%% \begin{align}
%%   \frac{\Gamma(x)}{\Gamma(n+x)}
%%   \geq
%%   \frac{\Gamma(\hat{x}) \exp{((\hat{x}-x)b})}{\Gamma(n+\hat{x})},
%% \end{align}
%% with
%% \begin{align}
%%   b=\Psi(n+\hat{x}) - \Psi(\hat{x}),
%% \end{align}
%% where, $\Psi(x)$ denotes the digamma function.

%% For $n \geq 1$,
%% \begin{align}
%% \frac{\Gamma(n+x)}{\Gamma(x)} \geq cx^{a},
%% \end{align}
%% with
%% \begin{align}
%%   a = (\Psi(n+\hat{x}) - \Psi(\hat{x}))\hat{x}, \\
%%   c = \frac{\Gamma(n+\hat{x})}{\Gamma(\hat{x})}\hat{x}^{-a}.
%% \end{align}

% The \nocite command causes all entries in a bibliography to be printed out
% whether or not they are actually referenced in the text. This is appropriate
% for the sample file to show the different styles of references, but authors
% most likely will not want to use it.
% \nocite{*}

\bibliography{network}% Produces the bibliography via BibTeX.

\end{document}